\documentclass{aa}  

\usepackage{graphicx}
\usepackage{txfonts}

\newcommand{\abs}[1]{\left\lvert#1\right\rvert}
\def\d{\displaystyle}

\begin{document}

   \title{Mixed properties of magnetohydrodynamic waves undergoing resonant absorption in the cusp continuum}

   \author{M. Goossens \inst{1}
          \and
          S.-X. Chen \inst{2}
          \and
          M. Geeraerts \inst{1}
          \and
          B. Li \inst{2}
          \and
          T. Van Doorsselaere \inst{1}
          }

   \institute{Centre for mathematical Plasma Astrophysics (CmPA), KU Leuven, Celestijnenlaan 200B bus 2400, B-3001 Leuven, Belgium\\
         \and
             Shandong Provincial Key Laboratory of Optical Astronomy and Solar-Terrestrial Environment, Institute of Space Sciences, Shandong University, Weihai 264209, People's Republic of China\\
             }

   \date{}

% \abstract{}{}{}{}{} 
% 5 {} token are mandatory
 
  \abstract
  % leave it empty if necessary  
   {Observations of magnetohydrodynamic (MHD) waves in the structured solar atmosphere have shown that these waves are damped and can thus contribute to atmospheric heating. In this paper, we focus on the damping mechanism of resonant absorption in the cusp continuum. This process takes places when waves travel through an inhomogeneous plasma.}
   {Our aim is to determine the properties of MHD waves undergoing resonant absorption in the cusp continuum in the transition layer 
 of a cylindrical solar atmospheric structure, such as a photospheric pore or a coronal loop. Depending on which quantities dominate, one can assess what type of classical MHD wave the modes in question resemble most.}
   {In order to study the properties of these waves, we analytically determine the spatial profiles of compression, displacement, and vorticity for waves with frequencies in the cusp continuum, which undergo resonant absorption. We confirm these analytical derivations via numerical calculations of the profiles in the resistive MHD framework.}
   {We show that the dominant quantities for the modes in the cusp continuum are the displacement parallel to the background magnetic field and the vorticity component in the azimuthal direction (i.e. perpendicular to the background magnetic field and along the loop boundary).}
   {}


   \maketitle
%
%-------------------------------------------------------------------

\section{Introduction}
 
In non-uniform plasmas, magnetohydrodynamic (MHD) waves behave differently than their counterparts in uniform plasmas of infinite extent \citep{GoossensEtAl2019}.  In the latter case, the MHD waves can be separated into Alfv\'{e}n waves and magnetosonic waves.  The situation is different in a non-uniform plasma.  The clear division between Alfv\'{e}n waves and magnetosonic waves is no longer present. The MHD waves have mixed properties.  The general rule is that MHD waves propagate both parallel vorticity, as in classic Alfv\'{e}n waves, and compression, as in classic magnetosonic waves. The present paper is  a companion to the paper by \citet{GoossensEtAl2020}, which dealt with resonant damping in the Alfv\'{e}n continuum.  In that paper, plasma pressure was neglected but here it plays an essential role. \citet{GoossensEtAl2020}  focused  on the properties of MHD waves that undergo resonant absorption in the Alfv\'{e}n continuum.  They  considered a straight magnetic field and in addition assumed  that the plasma is pressureless. This assumption removes the slow magnetosonic part of the spectrum as well as the resonant absorption in the cusp continuum. Since the focus here will be on the behaviour in the cusp continuum, we include plasma pressure in the analysis. 

Interest in the resonant behaviour of MHD waves has mainly been concentrated on resonant absorption in the Alfv\'{e}n continuum \citep{GoossensEtAl1992, GoossensEtAl2002a, SolerEtAl2013}. Recently, there has been increased interest in the cusp continuum because of the observation of MHD waves in the lower solar atmosphere. These waves have been observed in the chromosphere \citep{DePontieuEtAl2007, MortonEtAl2012, Verth&Jess2016} as well as in the photosphere \citep{DorotovicEtAl2008, Fujimura&Tsuneta2009, GrantEtAl2015, MoreelsEtAl2015, KeysEtAl2018, Gilchrist-MillarEtAl2020}. The oscillations observed in a photospheric magnetic pore  by, for example, \citet{GrantEtAl2015} display characteristics of rapidly decaying slow surface sausage modes (SSSMs). \citet{YuEtAl2017} analytically calculated the damping rate of these modes by resonant absorption in the slow continuum in the thin boundary limit. In a numerical study, \citet{ChenEtAl2018} compared the damping rates of SSSMs due to resonant absorption in the slow continuum with the damping rate due to Ohmic resistivity. Their numerical results were later confirmed by the analytical derivations in \citet{GeeraertsEtAl2020}. In another numerical study, \citet{ChenEtAl2020} compared the damping rates of slow surface kink modes (SSKMs) due to resonant absorption in the Alfv\'en and slow continua with the damping rate due to Ohmic resistivity. 

In the present investigation, we focus on  the eigenvalue problem (EVP) and try to understand what happens when the wave is actually damped in non-stationary MHD. We take the frequency to be complex and relate the spatial behaviour to the damping properties of the MHD wave. The temporal and spatial variation of the perturbed quantities is therefore described by the factor 

\begin{equation*}
\exp\{i (m \varphi +  k_z z - \omega t)\} \text{.}
\end{equation*}
Here, $m$ and $k_z$ are, respectively, the azimuthal and longitudinal wave numbers, and $\omega$ is the complex frequency with

\begin{equation*}
\omega = \omega_R + i \omega_I \text{, } \hspace{1cm} \omega_I = \gamma \text{.}
\end{equation*}
Since we focus on the EVP, we do not have to worry about possible  Gaussian damping that can occur in the temporal  evolution after the initial excitation. This Gaussian damping is well known for propagating \citep{PascoeEtAl2010, PascoeEtAl2011, PascoeEtAl2012, HoodEtAl2013} and standing kink waves \citep{Ruderman&Terradas2013, GuoEtAl2020} in coronal loops.  However,  it is not clear to us whether this behaviour has  been observed, in real life or  in numerical experiments, regarding the temporal evolution for slow waves in the photosphere.

Studies on resonant absorption have mainly focused on  the components of the displacement, the amount of absorbed energy,  and the damping rate. Little to no attention has been paid to the change in the spatial behaviour of fundamental quantities such as compression and vorticity.  Analytical solutions for the components of the Lagrangian displacement in the dissipative layer have been derived  by, for example, \citet{SakuraiEtAl1991} and \citet{GoossensEtAl1995} for resonant MHD waves in ideal and dissipative stationary MHD.

%\section{Resonant absorption} 
Resonant absorption has a long history in fusion plasma physics, space plasma physics, solar physics, and astrophysics.  A characterization was given by \citet{Parker1991}, who noted that resonant absorption in the Alfv\'{e}n continuum is to be expected when a wave with a phase velocity $\omega / k$ spans a region in which the variation of the Alfv\'{e}n velocity $v_A$ across the region provides the  resonance condition  $\omega / k = v_A$. Non-uniformity is key  to the process.  The Alfv\'{e}n velocity $v_A$  and  the Alfv\'{e}n frequency $\omega_A$ depend on position.  The resonance occurs at the position $r_A$ where the frequency of the wave $\omega$ is equal to the local Alfv\'{e}n frequency,  $\omega = \omega_A(r_A)$.

Resonant absorption has been studied extensively for frequencies in the Alfv\'{e}n continuum. For example,
since 2002 \citep{Ruderman&Roberts2002, GoossensEtAl2002a}, resonant absorption of kink waves has been a popular mechanism for explaining the rapid damping of standing and propagating MHD waves in coronal loops. The simple model invokes MHD waves superimposed on a cylindrical plasma column in static  equilibrium. Cylindrical coordinates $(r, \varphi, z)$ are used. The inhomogeneity necessary for resonant absorption to operate is usually provided by the equilibrium density $\rho_0(r)$, which varies  from  $\rho_i$ to $\rho_e$ in the interval $[R - l/2, \;\;R + l/2]$.  The density $\rho_0$ is constant in the internal and external parts of the loop with values $\rho_i$ and $\rho_e$

Relatively simple analytic expressions for the ratio of the damping time over the period for standing MHD waves can be obtained by using  the  thin tube and thin boundary (TTTB)  approximation. The TTTB approximation means 
that  $k_z R  \ll 1$ and $\; l/R  \ll 1$. Analytical expressions for the damping time $\tau_D$  were derived by \citet{Hollweg1990a}, \citet{GoossensEtAl1995}, \citet{Ruderman&Roberts2002}, \citet{VanDoorsselaereEtAl2004}, and \citet{Arregui&Goossens2019}. The expression given by \citet{Arregui&Goossens2019} is

\begin{equation}
 \frac{\tau_D}{\mbox{Period}} = \frac{ 1}{\d \abs{m}} \;\frac{4}{\pi^2}\; \frac{\alpha}{l/R} 
\;\frac{\rho_i + \rho_e}{\rho_i -\rho_e} \text{,}
\label{TauAG}
\end{equation}
where $\mbox{period} =  2 \pi / \omega_R$,$\;\gamma $ is the damping decrement, and $\tau_d = 1/\mid \gamma \mid  $  is the damping time. The  variation of density is confined to a layer of thickness $l$ and has steepness $\alpha.$

Resonant absorption in the cusp continuum has already been studied in previous works, such as \citet{Keppens1995}, \citet{SolerEtAl2009c},\citet{YuEtAl2017}, \citet{ChenEtAl2018}, and \citet{Yu&VanDoorsselaere2019}. For a straight field, there is no resonant damping of axisymmetric modes in the Alfv\'{e}n continuuum \citep{SakuraiEtAl1991}. Resonant damping in the cusp continuum occurs  for  axisymmetric ($m=0$) and non-axisymmetric motions if their frequency is in the cusp continuum. In this process, the wave excites local oscillations with the same frequency and with properties close to those of the slow waves from an infinite homogeneous plasma, as will be discussed in this paper. An analytical expression for the ratio of the damping time over the period is rather involved for axisymmetric motions and has been derived in \citet{YuEtAl2017}. An expression for non-axisymmetric motions can be found in \citet{SolerEtAl2009c}, where they compare the efficiency of the damping of the kink mode by resonant absorption in the cusp continuum with the resonant absorption in the Alfv\'en continuum. \citet{ChenEtAl2018} compare the damping of SSSMs in photospheric pore conditions by resonant absorption with that by electrical resistivity.

As explained, for example, in \citet{GoossensEtAl2019}, whether MHD waves in a given inhomogeneous plasma have Alfv\'enic, fast, or slow magnetosonic properties can be assessed by looking at the displacement components, as well as the compression and the vorticity carried by these waves. In this paper, our interest therefore lies in deriving the spatial profiles of the compression as well as the displacement and vorticity components of MHD waves with frequencies in the cusp continuum.

\section{ MHD waves with mixed properties for a straight field}

The mixed properties of the MHD waves in an inhomogeneous plasma arise because the equations that rule them are coupled.\ This was pointed out by, for example, \citet{GoossensEtAl2019}, who showed that the waves have features of both Alfv\'en waves (namely a non-zero parallel vorticity) and magnetosonic waves (non-zero compression). The coupling of the MHD equations is controlled by the  coupling functions  $ C_A$ and $C_S$, which were first introduced by \cite{SakuraiEtAl1991}. For a straight field,  $\vec{B_0}= B_z(r) \vec{1}_z$, these two constants take the simple form

\begin{equation}
C_A =  g_B P' = \frac{m}{r} B_z P', \;\; C_S = P' \text{.}
\label{CASStraightF}
\end{equation}
This means that the equations are  coupled because of $P'$. The importance of $P'$ for resonant absorption has been discussed by, for example, \citet{Hasegawa&Uberoi1982}, \citet{Hollweg&Yang1988}, \citet{Tirry&Goossens1995}, and \citet{SolerEtAl2013}.  

Since we are dealing with non-stationary MHD waves, the frequency $\omega$  is a complex quantity with an imaginary part expressing the damping rate. \citet{GoossensEtAl2020} studied the behaviour of MHD waves that undergo resonant absorption in the Alfv\'en continuum. They considered a pressureless plasma, which helps to simplify the analysis and is often a good approximation of reality for the Alfv\'en resonance. The situation for the slow resonance is different, however. First of all, plasma pressure is essential because there are no slow waves in a pressureless plasma. Secondly, there is resonant absorption in the cusp continuum for axisymmetric waves ($m=0$) in the same way as for non-axisymmetric waves.

The expressions for the components of the displacement $\vec{\xi}$ and compression can be found, for example, in \citet{GoossensEtAl2019} (see equation (45) therein):

\begin{eqnarray}
\xi_r & = & \frac{1}{\rho_0 (\omega^2 - \omega_A^2)} \frac{d P'}{dr}, \nonumber  \\
&& \nonumber \\ 
\xi_{\perp} & = & \xi_{\varphi} \;\; = \;\; \frac{ i m/r}{\rho_0 (\omega^2 - \omega_A^2)} P', \nonumber  \\
&& \nonumber  \\
\xi_{\parallel} & = & \xi_z  \;\; = \;\; i k_z \frac{v_s^2}{v_s^2 + v_A^2}\; \frac{1}{ \rho_0 (\omega^2 - \omega_C^2) } P' , \nonumber \\
&& \nonumber  \\
\nabla \cdot \vec{\xi} & = & \frac{-\omega^2 }  {(v_s^2 + v_A^2)\;} \frac{1}{  \rho_0  \;(\omega^2 - \omega_C^2)} \;  P'  .
\label{EqStrFieldA}      
\end{eqnarray}
The quantities  $\omega_A$ and  $\omega_C$  are the local Alfv\'{e}n frequency and  the local cusp frequency, respectively. They are  defined as 

\begin{eqnarray}
\omega_A^2  & =  & \frac{\d (\vec{k} \cdot
\vec{B})^2}{\d \mu \rho} \;\;=\;\; k_z^2 v_A^2 \;\; = \;\; k_{\parallel} v_A^2, \nonumber \\
&& \nonumber \\ 
\omega_C^2  & =  & \frac{v_s^2}{v_s^2 + v_A^2}  \omega_A^2   \;\; = \;\; \frac{v_s^2}{v_s^2 + v_A^2} k_z^2 v_A^2 \;\; = \;\; k_z^2 v_{C}^2, 
\label{CuspFre1}
\end{eqnarray}
where $v_A$, $v_s$, and $v_C$ are the Alfv\'{e}n speed,  sound speed, and cusp speed, respectively. They are defined as 
\begin{equation}
 v_A^2 = \frac{\d B_0^2}{\d \mu \rho_0},\;\;  v_s^2 = \frac{\d \gamma  p_0}{\d  \rho_0}, \;\; v_C^2 = \frac{\d v_A^2 v_s^2}{\d  v_A^2 + v_s^2},
\label{VAVS}
\end{equation}
with $\gamma = c_p/c_v$ the ratio of the specific heats and $p_0$ the equilibrium thermal pressure. This should not be confused with the temporal decrement in equation (\ref{TauAG}). In a non-uniform plasma, the characteristic frequencies $\omega_A$ and $\omega_C$ depend on position and define the Alfv\'{e}n and cusp continua.

Similarly,  equation (53) from Goossens et al. (2019) gives the expressions for the components of $\nabla \times \vec{\xi}$:

\begin{eqnarray}
(\nabla \times \xi)_r & = &   k_z  \frac{m}{r} \frac{v_A^2}{v_s^2 + v_A^2} \frac{\omega^2}{\rho_0 (\omega^2 - \omega_A^2)(\omega^2 - \omega_C^2)}P' \nonumber \\
&& \nonumber \\
(\nabla \times \xi)_{\varphi}  & = & (\nabla \times \xi)_{\perp} \nonumber \\
 & = &   i k_z \frac{\d 1} {\d \left \{\rho_0 (\omega^2 - \omega_C^2) \right \} ^2}
\frac{\d d}{\d dr} \left \{\rho_0 (\omega^2 - \omega_C^2)\right \}   \frac{\d v_s^2}{\d v_A^2 + v_s^2} P' \nonumber  \\
& + & i k_z \frac{\d \omega^2}{\d \rho_0 (\omega^2 - \omega_A^2)(\omega^2 - \omega_C^2)}\frac{\d v_A^2}{\d v_A^2 + v_s^2} \frac{\d d P'}{\d dr}
\nonumber \\
&& \nonumber \\
 (\nabla \times \xi)_z & =  & (\nabla \times \xi)_{\parallel} \nonumber  \\
& = & -  i \frac{m}{r} \frac{\d 1} {\d \left \{\rho_0 (\omega^2 - \omega_A^2) \right \} ^2}
\frac{\d d}{\d dr} \left \{\rho_0 (\omega^2 - \omega_A^2)\right \}  P'
\label{EqStrFieldB}
.\end{eqnarray}
Equations \eqref{EqStrFieldA} and \eqref{EqStrFieldB} clearly show that $P'$ plays the role of a coupling function. The horizontal components of vorticity $(\nabla \times \vec{\xi})_r $ and $(\nabla \times \vec{\xi})_{\varphi}$ are always non-zero, whereas the parallel vorticity  $(\nabla \times \vec{\xi})_z$ is only non-zero if

\begin{equation}
\frac{\d d}{\d dr} \left \{\rho_0 (\omega^2 - \omega_A^2)\right \} \neq 0
\label{NonUniformOmegaA}
,\end{equation}
that is to say, in the inhomogeneous transition layer (TL) at the loop boundary (where the density varies with $r$). In a non-uniform plasma, all the quantities are non-zero. As a consequence, the MHD waves are neither pure Alfv\'en waves (which have zero compression) nor pure magnetosonic waves (which have zero parallel vorticity). As was pointed out by \citet{GoossensEtAl2002a, GoossensEtAl2006, GoossensEtAl2008, GoossensEtAl2011, GoossensEtAl2012, GoossensEtAl2019}, MHD waves in a non-uniform plasma have mixed properties that change as the waves travel through the inhomogeneous part of the plasma.

\citet{GoossensEtAl2020} focused on MHD waves that undergo damping in the Alfv\'{e}n continuum. They showed that $(\nabla \times \vec{\xi})_{\parallel}$ is much larger than $(\nabla \times \vec{\xi})_r $ and $(\nabla \times \vec{\xi})_{\varphi}$ in the inhomogeneous TL. Since in a homogeneous plasma of infinite extent only Alfv\'en waves propagate parallel vorticity, this showed that waves undergoing resonant absorption in the Alfv\'en continuum are very close to Alfv\'en waves in their nature despite having non-zero compression.

Here we study MHD waves that undergo damping by resonant absorption in the cusp continuum. In what follows, we show that, contrary to the Alfv\'en continuum, the parallel component of vorticity does not play any significant role here. Instead, we will show that the perpendicular component of the vorticity, $(\nabla \times \vec{\xi})_{\varphi}$, is dominant. In particular, the first term on the right-hand side of  $(\nabla \times \xi)_{\perp} $ will be important. This term is non-zero when

\begin{equation}
\frac{\d d}{\d dr} \left \{\rho_0 (\omega^2 - \omega_C^2)\right \} \neq 0
\label{NonUniformOmegaC}
.\end{equation}

In the particular case of a constant straight field, $\vec{B}_0 = B_0 \vec{1}_z $,  the equation of motion demands that the thermal pressure $p_0$ also be constant. We then have the simplifying relations

\begin{eqnarray}
\rho_0 v_A^2 =  \frac{B_0^2}{\mu_0} = \mbox{constant},\;\;\;
\rho_0 \omega_A^2 = k_z ^2 \frac{B_0^2}{\mu_0} = \mbox{constant}, \nonumber \\
\frac{\d v_s^2}{\d v_A^2 + v_s^2}  = 
\mbox{constant}, \;\;\;  \rho_0 \omega_C^2 =  \rho_0 \frac{v_s^2}{v_s^2 + v_A^2} \omega_A^2 = \mbox{constant},
\nonumber \\
\label{ConstantAC}
\end{eqnarray}
and the spatial variation of $\omega_A^2$  and  $\omega_C^2$ is then solely due to the spatial variation of the equilibrium density $\rho_0$. Hence, 

\begin{equation}
\frac{\d d\rho_0}{\d dr} \neq 0
\label{NonDensity}
\end{equation}
is the important quantity for the resonant behaviour both in the Alfv\'{e}n continuum and in the cusp continuum in case of a constant background magnetic field. For the cusp continuum, we also need $v_s^2  \neq 0$. We note that we will not use the simplifying assumption that the straight magnetic field $B_0 \vec{1}_z$ is constant because $p_0$ and $B_0$ are not constant in pores. The simplifications in Eq. \eqref{ConstantAC} are thus not used in what follows.

The expression for the parallel component of vorticity under this assumption was given by equation (11) in \citet{GoossensEtAl2020}:

\begin{eqnarray}
(\nabla \times \vec{\xi})_{\parallel} &=& (\nabla \times \xi)_z \nonumber\\
 &=& -  i \frac{m}{r} \frac{\d  \omega^2 } {\d \left \{\rho_0 (\omega^2 - \omega_A^2) \right \} ^2} \; \frac{\d d  \rho_0}{\d dr} \; P'.
\label{vorticityZ1}
\end{eqnarray}
The perpendicular component can similarly be calculated by noting, using Eq. \eqref{ConstantAC}, that

\begin{eqnarray*}
\frac{\d d}{\d dr} \left \{\rho_0 (\omega^2 - \omega_C^2)\right \}  & = & \omega^2 \; \frac{\d d \rho_0}{\d dr}
\end{eqnarray*}
to find that

\begin{eqnarray}
 (\nabla \times \vec{\xi})_{\perp} &=& (\nabla \times \xi)_{\varphi} \nonumber \\ &=&  
 i k_z  \frac{v_s^2}{v_s^2 + v_A^2} \frac{\d  \omega^2 } {\d \left \{\rho_0 (\omega^2 - \omega_C^2) \right \} ^2}\; \frac{\d d  \rho_0}{\d dr} \; P'
  \nonumber  \\
 & +& i k_z \frac{\d \omega^2}{\d  (\omega^2 - \omega_A^2)} \;
\frac{\d v_A^2}{\d v_A^2 + v_s^2}  \;
\frac{\d 1}{\d \rho_0 (\omega^2 - \omega_C^2)}
\frac{\d d P'}{\d dr}.
\label{vorticityPerp1}
\end{eqnarray}
The first term on the right-hand hand side of $(\nabla \times \xi)_{\varphi} $  is the dominant term for frequencies in the cusp continuum,  as we shall explain in the next section.  It is  remarkably similar to Eq. (\ref{vorticityZ1}). 

We recall from Eq. \eqref{EqStrFieldA} that $P'$ is related to $\nabla \cdot \vec{\xi}$ by an algebraic equation. Hence, it is straightforward to express $(\nabla \times \vec{\xi})_{\parallel}$ and $(\nabla \times \vec{\xi})_{\perp}$ in terms of $\nabla \cdot \vec{\xi}$, so that parallel and perpendicular vorticity go together with compression in a non-uniform plasma.

\section{Analysis of slow resonant waves  close to the resonant point} \label{analysis}

\vspace{2mm} \noindent We now consider a standing MHD wave that is damped by undergoing resonant absorption in the cusp continuum. Therefore, the longitudinal wavenumber $k_z$ is real and the frequency is complex:

\begin{equation}
\omega = \omega_R + i \gamma, \;\;  \gamma = \omega_I, \;\;\omega_R = \omega_C(r_C)
\label{Complexfrequency}
.\end{equation}
The frequency $\omega_R$ and the damping decrement $\gamma$ are related to the period and the damping time as
\begin{eqnarray}
\frac{1}{\abs{\gamma}} = \tau_D, \;\;\; \mbox{Period} = \frac{2 \pi}{\omega_R}, \nonumber  \\ 
\frac{\tau_D}{\mbox{Period}} = \frac{1/\abs{\gamma}} {2 \pi / \omega_R} = \frac{1}{2 \pi} \frac{\omega_R}{\abs{\gamma}}, \nonumber \\
\frac{\abs{\gamma}}{\omega_R} = \frac{1}{2 \pi} \frac{\mbox{Period}}{\tau_D}.
\label{PeriodDampingTime}
\end{eqnarray}

In view of the periods and damping times observed in waves in the solar corona, it is a good approximation to assume

\begin{equation}\label{weakDamping}
\frac{\abs{\gamma}}{\omega_R} \ll 1 \text{,}
\end{equation}
which represents weak damping. The focus goes to the position $r_C$, where the ideal resonance occurs: $\omega^2 = \omega_C^2 (r_C)$. The comments in \citet{GoossensEtAl2020} about the position where the resonance occurs, explaining that it is actually off the real axis, apply here as well. As explained there, the assumption of non-stationary MHD (namely that $\omega$ is complex) entails that the resonant position $r_C$ will be complex as well. In the case of a weak damping, as expressed by Eq. \eqref{weakDamping}, the imaginary part of $r_C$ will also be small compared to its real part, and we can make the approximation to neglect it and consider $r_C$ to be real.

The results that are discussed in this section actually describe the MHD waves just outside the infinitely thin resonant layer. The waves inside this layer are not addressed, as resistive effects become too important to be ignored there, so that ideal MHD is not a valid approximation. However, barring the resonant layer, the ideal MHD solutions are a good approximation of the full resistive solution for sufficiently small damping, and therefore the results from the ideal MHD calculations that are made here make physical sense.

In what follows, we simplify the expressions for the wave variables by retaining only the dominant terms in $\abs{\gamma} / \omega_R$. The dominant contributions to $\xi_{\parallel} = \xi_z$, $\nabla \cdot \vec{\xi}$, $(\nabla \times \xi)_r$, and $(\nabla \times \xi)_{\varphi}$ are caused by the terms

\begin{equation*}
\frac{1}{\rho_0 (\omega^2 - \omega_C^2)}  \;\; \text{and} \;\; \frac{1}{\left \{\rho_0 (\omega^2 - \omega_C^2)\right\}^2} \text{,}
\end{equation*}
which lead, respectively, to behaviour proportional to

\begin{equation*}
 \frac{  \tau_D }{ \mbox{Period}} \;\; \text{and} \;\; \left\{ \frac{  \tau_D }{ \mbox{Period}}\right\}^2 \text{,}
\end{equation*}
as we will derive next. The quantities $\xi_r$, $\xi_{\perp} = \xi_{\varphi}$, and $(\nabla \times \xi)_z$ do not have a factor of the form 

\begin{equation} \label{n}
\frac{1}{\left \{\rho_0 (\omega^2 - \omega_C^2)\right\}^n}, n \geq 1
\end{equation}
and are therefore independent of $\tau_D /  \mbox{Period}$. These quantities do not feel the cusp resonance and thus
do not play an important role in the process of resonant absorption in the cusp continuum.

We will first consider the parallel component of displacement $\xi_z = \xi_{\parallel}$ and compression $\nabla \cdot \vec{\xi}$. These two quantities are proportional to Eq. \eqref{n} with $n = 1$. We use the following intermediate results for weak damping:

\begin{eqnarray}
\omega^2 & = & ( \omega_R + i \gamma)^2  \approx \omega_R ^2 + 2 i \gamma \omega_R, \nonumber\\
\rho_0 (\omega^2 - \omega_C^2 )  & \approx &  2 i \rho_0(r_C) \omega_R^2
\;\frac{\gamma}{\omega_R}, \nonumber \\
\frac{1}{\rho_0 (\omega^2 - \omega_C^2)} & \approx & - i \frac{\pi} {\rho_0(r_C) \omega_R^2} 
\frac{  \tau_D }{ \mbox{Period}}. \label{o-o_C}
\end{eqnarray}
We then find the following approximations for parallel displacement and compression:

\begin{eqnarray}
\xi_z &\approx &  k_z\; \frac{v_s^2}{v_s^2 + v_A^2}\; \frac{\pi} {\rho_0  \; \omega_R^2}  \;
\; \frac{ \tau_D }{ \mbox{Period}}\; P',  \;\nonumber \\
\nabla \cdot \vec{\xi} & \approx  & - i \frac{\pi} {\rho_0  \;(v_s^2  + v_A^2) }  \;
\frac{  \tau_D }{ \mbox{Period}}  P'.
\label{ParallelDCompression}
\end{eqnarray}

We will now consider the radial component of vorticity $(\nabla \times \xi)_r$ in Eq. \eqref{EqStrFieldB}. Like $\xi_z$ and $\nabla \cdot \vec{\xi} $,  this quantity is proportional to Eq. \eqref{n} with $n=1$. In addition, there is a factor $1/(\omega^2 - \omega_A^2)$. Since

\begin{equation}\label{o-o_A}
\omega^2 - \omega_A^2 \;\;  \approx \;\;  \omega_C^2 -  \omega_A^2  \;\;
\approx \;\; - \frac{v_A^2}{v_s^2} \omega_R^2 \text{,} \\
\end{equation}
it follows that 

\begin{eqnarray}
(\nabla \times \xi)_r & \approx  & - i \; \pi \; k_z \; \frac{m}{r}\; \frac{v_s^2}{v_s^2 + v_A^2}\; 
\frac{1 }{\rho_0 \; \omega_R^2} \;
\frac{ \tau_D }{ \mbox{Period}} \;P'  
\label{VorticityRadial}
.\end{eqnarray}

We now turn to the perpendicular component of vorticity, $(\nabla \times \vec{\xi})_{\varphi} $. The first term in its expression from Eq. \eqref{EqStrFieldB} is the dominant term since it contains the factor Eq. \eqref{n} with $n=2$. We thus discard the second term and find,
from Eq. \eqref{o-o_C}, that the dominant factor is approximated by 

\begin{equation*}
\frac{1}{ \left \{\rho_0 (\omega^2 - \omega_C^2) \right \} ^2} \;\approx \;- \frac{\pi ^2 }{\rho_0^2 \; \omega_R^4} \;
\left\{ \frac{  \tau_D }{ \mbox{Period}}\right\}^2 \text{,}
\end{equation*}
while for the factor with the derivative we have

\begin{eqnarray*}
\frac{\d d}{\d dr} \left \{\rho_0 (\omega^2 - \omega_C^2)\right \} & = & \frac{\d d \rho_0}{\d dr} \left( \omega^2 - \omega_C^2 \right)  \; -  \; \rho_0 k_z^2 \frac{\d d v_{C}^2}{\d dr}\\
& \approx & \frac{\d d \rho_0}{\d dr} \frac{i \omega_R^2}{\pi} \frac{\mbox{Period}}{\tau_D}  \; -  \; \rho_0 k_z^2 \frac{\d d v_{C}^2}{\d dr}\\
& \approx & -  \; \rho_0 k_z^2 \frac{\d d v_{C}^2}{\d dr}
\end{eqnarray*}
since $\tau_D/\text{Period} \ll 1$.

\noindent Hence, the result for the perpendicular vorticity is given by

\begin{eqnarray}
(\nabla \times \xi)_{\varphi}  & \approx &   
  i k_z^3 \frac{\pi^2}{\rho_0 \omega_R^4} \frac{\d v_s^2}{\d v_A^2 + v_s^2} \nonumber \frac{\d d v_C^2}{\d dr}
\left\{ \frac{  \tau_D }{ \mbox{Period}}\right\}^2  P'.
 \label{VorticityPerp}
 \end{eqnarray}

Next, we will simplify the expression in Eq. \eqref{EqStrFieldB} for the parallel component of vorticity, $(\nabla \times \vec{\xi})_z $. Since there are no terms that contain $\omega^2 - \omega_C^2$,  we can again forget about the damping decrement according to Eq. \eqref{weakDamping} and use 

\begin{equation*}
 \omega^2  \approx  \omega_R^2  =  \omega_C^2(r_C) \text{.}
\end{equation*}
From Eq. \eqref{o-o_A}, we have the following:

\begin{eqnarray*}
\frac{\d d}{\d dr} \left \{ \rho_0 (\omega^2 - \omega_A^2) \right \} & = &
\frac{\d d \rho_0}{\d dr} \left( \omega^2 - \omega_A^2 \right)  \; - \; \rho_0 k_z^2 \frac{\d d v_{A}^2}{\d dr}\\
& \approx &   - \frac{\d d \rho_0}{\d dr} \frac{v_A^2}{v_s^2} \omega_R^2 \; - \; \rho_0 k_z^2 \frac{\d d v_{A}^2}{\d dr}
.\end{eqnarray*}
Then, from this and Eq. \eqref{o-o_A}, we find 

\begin{eqnarray}
 (\nabla \times \xi)_z 
& \approx  & - i \frac{m}{r}  \left(\frac{v_s ^2}{v_A^2} \right)^2  \frac{1}{\rho_0^2 \omega_R^4} \nonumber \\
& \cdot & \left\{ -\frac{\d d \rho_0}{dr} \left(\frac{v_A ^2}{v_s^2} \right) \omega_R^2 - \rho_0 k_z^2 \frac{\d d v_A^2}{\d dr} \right\}  P'.
 \label{VorticityPar}
\end{eqnarray}

Finally, from Eq. \eqref{EqStrFieldA}, the radial and perpendicular components of the displacement are found to be

\begin{eqnarray}
\xi_r & \approx &  -\frac{1}{\rho_0}\; \frac{v_s^2}{ v_A^2}\; \frac{1}{ \omega_R^2}  \frac{\d d P'}{\d dr}, \nonumber  \\
&& \nonumber \\ 
\xi_{\perp} & = & \xi_{\varphi}  \;\; \approx \;\; - i \frac{ m}{r}\;\frac{1}{\rho_0}\; \frac{v_s^2}{ v_A^2}\; \frac{1}{ \omega_R^2}  P'.
\label{CompRP}      
\end{eqnarray}

The case of axisymmetric ($m=0$) slow waves is a bit simpler because some quantities become $0$. We note that $\xi_{\varphi},\; (\nabla \times \xi)_r, $ and$ \;(\nabla \times \xi)_{z}  $ are indeed proportional to $m$ and thus vanish for the $m=0$ modes. On the other hand, the expressions for $\xi_r$, $\xi_z$, $\nabla \cdot \vec{\xi}$, and $(\nabla \times \xi)_{\varphi}$ do not depend explicitly on $m$. However, they do depend implicitly on $m$ since $P'$ depends on $m$. In any case, the expressions for axisymmetric modes around the ideal cusp resonant position for the components of the Lagrangian displacement and for the compression are given by

\begin{eqnarray}
\xi_r & = &  -\frac{1}{\rho_0}\; \frac{v_s^2}{ v_A^2}\; \frac{1}{ \omega_R^2}  \frac{\d d P'}{ \d dr}, \nonumber  \\
&& \nonumber \\ 
\xi_{\perp} & = &\xi_{\varphi} \;\; = \;\; 0, \nonumber \\
\xi_z &\approx &  k_z\; \frac{v_s^2}{v_s^2 + v_A^2}\; \frac{\pi} {\rho_0  \; \omega_R^2}  \;
\frac{ \tau_D }{ \mbox{Period}}\; P' , \;  \nonumber \\
\nabla \cdot \vec{\xi} & \approx  & - i \frac{\pi} {\rho_0  \;(v_s^2  + v_A^2) }  
\frac{  \tau_D }{ \mbox{Period}}  P'.
\label{AxiDisCom}
\end{eqnarray}
The components of the vorticity  are given by

\begin{eqnarray}
(\nabla \times \xi)_r & = & 0, \nonumber \\
(\nabla \times \xi)_{\varphi}  & \approx &   
 i k_z^3 \frac{\pi^2}{\rho_0 \omega_R^4} \frac{\d v_s^2}{\d v_A^2 + v_s^2} \nonumber \frac{\d d v_C^2}{\d dr}
\left\{ \frac{  \tau_D }{ \mbox{Period}}\right\}^2  P',
\nonumber \\
(\nabla \times \xi)_{z}   & = & 0. 
\label{Vorticity}
\end{eqnarray}

The conclusion of this section is that $\xi_z\;$, $\nabla \cdot \vec{\xi}$, and $(\nabla \times \xi)_{\varphi} $ are the dominant quantities and that
\begin{equation}
\vec{\xi}  \approx  \xi_z \vec{1}_z, 
\;\; \nabla \times \vec{\xi}   \approx  (\nabla \times \vec{\xi})_{\varphi}\;  \vec{1}_{\varphi} \text{.}
\label{DominantB}
\end{equation}
The waves in the cusp continuum are compressive with large parallel motions and carry a huge amount of perpendicular vorticity. This implies that the waves are similar to the slow magnetosonic waves from an infinite homogeneous plasma, although they also have some slight classical Alfv\'en wave properties since their parallel vorticity is non-zero. The results are in the first order independent of the wave number $m$.

\citet{GoossensEtAl2020} used the approximate analytical expression Eq. \eqref{TauAG} for $\tau_D/ \mbox{Period}$ for the damping in the Alfv\'{e}n continuum to obtain expressions that depend on the equilibrium quantities $\rho_i, \rho_e, l/R$, and $\alpha$ and on the wave numbers $m$ and $k_z R$. This has not been done in the present case since the analytical expressions involving $\tau_D/ \mbox{Period}$ derived in this section are rather involved.

\section{Spatial variation for resonant slow (and Alfv\'en) waves}

In this section, we verify our analytical predictions with numerical calculations. The aim is to understand the behaviour of waves in the cusp continuum by looking at the profiles of compression as well as at the displacement and vorticity components. Because of the inclusion of resistivity in the calculations, the quantities that have a pole at the resonant position in ideal MHD will actually be very large, though finite, thanks to dissipation. 

A brief description for the resistive solutions is as follows, and we refer interested readers to \citet{ChenEtAl2018} and \citet{ChenEtAl2020} for more details. We modelled a photospheric waveguide as a static, straight, field-aligned plasma cylinder. The equilibrium quantities depend only on $r$ in such a way that the equilibrium configuration comprises a uniform interior (denoted by the subscript i) and a uniform exterior (subscript e) as well as a TL continuously connecting the two. In the TL, the squared cusp ($v_{\rm C}^2$)  and adiabatic sound speeds ($v_{\rm S}^2$) were taken to depend on $r$ through $\sin\left[\pi (r-R)/l\right]$, where $R$ is the mean waveguide radius and $l$ is the TL width. With pores and sunspot umbrae in mind, we fixed the characteristic speeds $[v_{\rm Si},  v_{\rm Se}, v_{\rm Ae}]$  at $[0.5, 0.75, 0.25] v_{\rm Ai}$. We adopted a rather narrow TL by adopting an $l/R$ of $0.1$. Working in the framework of linear, resistive MHD,  we then formulated and numerically solved the relevant EVP with the general-purpose finite-element code PDE2D \citep{Sewell1988}. We examined the resonant damping of both SSKMs and SSSMs. This was done by locating the range of sufficiently large magnetic Reynolds numbers ($R_{\rm m}$)  where the eigenfrequencies depend only on the dimensionless axial wavenumber ($kR$) but are $R_{\rm m}$-independent. We refer to this range of $R_{\rm m}$  as 'the resonant regime'. It turns out that the critical $R_{\rm m}$, only beyond which the collective modes enter the resonant regime, is somehow dependent on $kR$. In the figures below, we consistently choose an $R_{\rm m}$ that is barely inside the resonant regime. 

Looking back at Eqs. \eqref{EqStrFieldA} and \eqref{EqStrFieldB} for the general expressions of compression as well as displacement and vorticity components, we expect each of these quantities to be non-zero in a uniform plasma, except the parallel vorticity. In a non-uniform plasma, parallel vorticity is also expected to be non-zero. Near the cusp resonant position, the perpendicular component of vorticity is expected to be much larger than its radial and parallel components, and the parallel component of displacement is expected to be much larger than its radial and perpendicular components.

\begin{figure}
   \centering
   \includegraphics[width=\hsize]{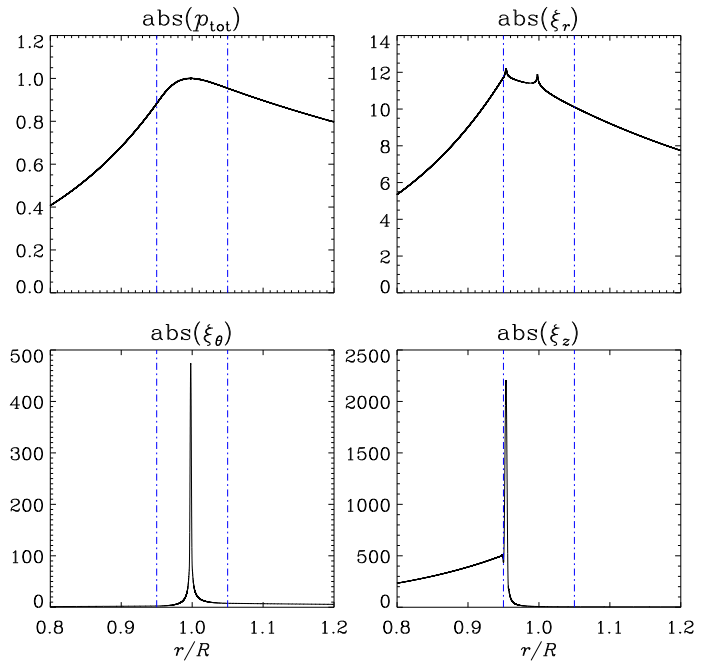}
      \caption{Modulus of the total pressure (top left), radial displacement (top right), perpendicular displacement (bottom left), and parallel displacement (bottom right) for the slow surface kink ($m=1$) mode. Here, $l/R = 0.1$, $k_z R =0.7$, and $R_m = 10^9$ have been taken.}
         \label{wl01_kR07eigen}
   \end{figure}

\begin{figure}
   \centering
   \includegraphics[width=\hsize]{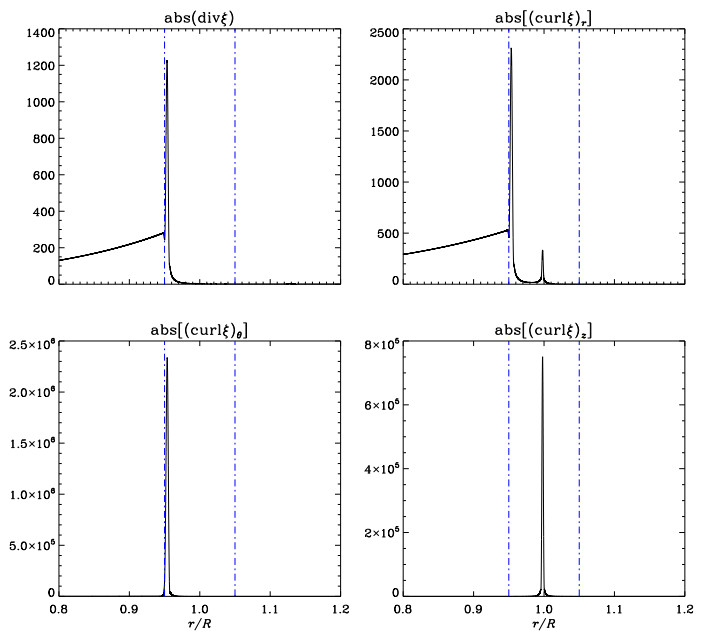}
      \caption{Modulus of compression (top left), radial vorticity (top right), perpendicular vorticity (bottom left), and parallel vorticity (bottom right) for the slow surface kink ($m=1$) mode. Here, $l/R = 0.1$, $k_z R =0.7$, and $R_m = 10^9$ have been taken.}
         \label{wl01_kR07}
   \end{figure}

We will first look at the SSKM. Figure \ref{wl01_kR07eigen} shows the total pressure and the three components of the Lagrangian displacement, whereas Figure \ref{wl01_kR07} shows compression and the three components of the vorticity of an SSKM with a frequency ranging in both the Alfv\'en and cusp continua. The cusp resonant position is close to the left border of the inhomogeneous layer at a position $r/R \approx 0.955$, while the Alfv\'en resonant position is situated near $r/R \approx 1$. 

As can be seen in the two mentioned figures, the parallel displacement ($\xi_z$), compression ($\nabla \cdot \vec{\xi}$), radial vorticity ($(\nabla \times \vec{\xi})_r$), and perpendicular vorticity ($(\nabla \times \vec{\xi})_{\varphi}$) peak at the cusp resonant position. This happens because of the factor \eqref{n} in their respective expressions, as is visible in Eqs. \eqref{EqStrFieldA} and \eqref{EqStrFieldB}, and the fact that $P'$ is constant at the cusp resonant position \citep{SakuraiEtAl1991}. However, only the perpendicular vorticity has this factor with $n=2$; the others have it with $n=1$. Therefore, perpendicular vorticity is expected to be the most dominant quantity in the cusp resonance; this is confirmed in the figures, which show that perpendicular vorticity has an amplitude three orders of magnitude higher than both the radial vorticity and compression. 

At the Alfv\'en resonant position, parallel vorticity and perpendicular displacement are the dominant quantities as they peak at the Alfv\'en resonant position. This was already discussed in \citet{GoossensEtAl2020} for a pressureless plasma. Radial and perpendicular displacement as well as radial and perpendicular vorticity and compression are then non-zero but definitely smaller than parallel vorticity and perpendicular displacement. It should be noted that the radial displacement jumps at both the cusp and Alfv\'en resonant positions, while the total pressure is continuous at both positions. This is in accordance with the derivations of \citet{SakuraiEtAl1991} in the case of a straight background magnetic field.

\begin{figure}
   \centering
   \includegraphics[width=\hsize]{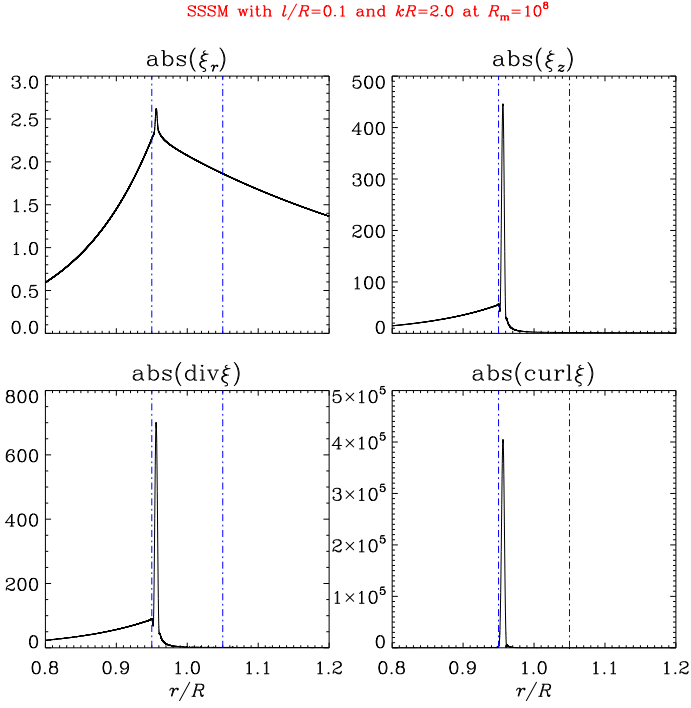}
      \caption{Modulus of the radial displacement (top left), parallel displacement (top right), compression (bottom left), and perpendicular vorticity (bottom right) for the slow surface sausage ($m=0$) mode. Here, $l/R = 0.1$, $k_z R =2$, and $R_m = 10^8$ have been taken.}
         \label{SSSM}
   \end{figure}

We will now take a look at the particular case of the SSSM (see Figure \ref{SSSM}). Since in this case $m=0$, equations \eqref{AxiDisCom} and \eqref{Vorticity} tell us that the perpendicular component of displacement as well as the radial and parallel components of vorticity are zero everywhere. Figure 3 shows that, again, the compression, perpendicular vorticity, and parallel displacements are the dominant quantities as they peak at the cusp resonant position near $r/R=0.955$. The perpendicular vorticity is again the most dominant. As pointed out by \citet{SakuraiEtAl1991}, there is no resonant absorption in the Alfv\'en continuum for the sausage modes. This explains the absence of a second peak near $r/R=1$.

The dissipation in the numerical calculation is physical  due to magnetic diffusivity $\eta$. It is instructive to try to understand the behaviour of the solution in the dissipative layer. The behaviour of the eigenfunctions in the dissipative layer in visco-resisitive non-stationary MHD was studied for incompressible plasmas by \citet{RudermanEtAl1995}. For incompressible plasmas, the cusp and Alfv\'{e}n resonances coincide. A simpler mathematical treatment was given by \citet{Tirry&Goossens1996}  for compressible plasmas. They treated the Alfv\'{e}n resonance in detail but did not give expressions for the cusp resonance. Mathematical aspects of the cusp resonance in resistive non-stationary MHD  were discussed in, for example, \citet{TirryEtAl1998} and \citet{PinterEtAl2007}. A further discussion on the eigenfunctions for the Alfv\'{e}n resonance can be found in \citet{GoossensEtAl2011}.

The important quantities in this discussion are $\delta_A$ and $\Lambda_A$ for the Alfv\'{e}n resonance and
$\delta_C$ and $\Lambda_C$ for the slow resonance. Expressions for $\delta_A$  and  $\delta_C$ (the thicknesses of the dissipative layers) are given in \citet{SakuraiEtAl1991}. When only magnetic diffusity is taken into account, 

\begin{equation*}
\delta_A \; = \;  \left\{  \frac{\omega  \eta}{\mid \Delta_A \mid} \;  \right \}^{1/3} \text{,} \hspace{0.7cm } \text{and} \hspace{0.7cm }\delta_C \; = \; \left\{  \frac{\omega  \eta}{\mid \Delta_C \mid} \;  \frac{\omega_C^2}{\omega_A^2} \right \}^{1/3} \text{,}
\end{equation*}
with  

\begin{equation*}
\Delta_A \; = \; \frac{d}{dr}( \omega^2   -\omega_A^2(r)) \text{,} \hspace{0.7cm } \text{and} \hspace{0.7cm }  \Delta_C \; = \; \frac{d}{dr}( \omega^2   -\omega_C^2(r)) \text{.}
\end{equation*}
The expression for $\Lambda_A$ is given in \citet{Tirry&Goossens1996}. The corresponding expression for $\Lambda_C$ can be found in \citet{TirryEtAl1998} and in \citet{PinterEtAl2007}. The $\Lambda_C$ is formally the same as $\Lambda_A$, with the index $A$ replaced by the index $C$:

\begin{equation*}
\Lambda_A \; = \; \frac{2 \omega_I \omega_A}{ \delta_A \abs{\Delta_A}} \text{,} \hspace{0.7cm } \text{and} \hspace{0.7cm } \Lambda_C \; = \; \frac{2 \omega_I \omega_C}{ \delta_C \abs{\Delta_C}} \text{.}
\end{equation*}

From the analysis in \citet{Tirry&Goossens1996}, it follows that the change from non-oscillatory behaviour to oscillatory behaviour occurs at about $\Lambda \approx 2$. For the SSKM in Figures \ref{wl01_kR07eigen}-\ref{wl01_kR07}, we used $ l/R = 0.1$, $R_m = 10^9$, and $k_z R = 0.7$. The  values for $\Lambda_A$ and $\Lambda_C$ are then

\begin{equation*}
\Lambda_A  = 0.029 \text{,} \hspace{0.7cm } \text{and} \hspace{0.7cm } \Lambda_C =0.41 \text{.}
\end{equation*}
For the SSSM in Figure \ref{SSSM}, we used $l/R = 0.1$, $R_m = 10^8$, and $k_z R = 2$, so that for $\Lambda_C$ we have

\begin{equation*}
 \Lambda_C = 0.73 \text{.}
\end{equation*}

These low $\Lambda_A$ and $\Lambda_C$ values confirm that the non-stationary resistive solutions are in the non-oscillatory regime, as can clearly be seen in Figures \ref{wl01_kR07eigen}-\ref{SSSM}.

\section{Conclusions}

In this paper, we studied the properties of MHD waves undergoing resonant absorption in the cusp continuum in the framework of non-stationary ideal MHD. As discussed by \citet{GoossensEtAl2019}, MHD waves in an inhomogeneous plasma do not have clearly distinguished properties that would permit one to classify them as Alfv\'en waves, fast magnetosonic waves, or slow magnetosonic waves, unlike in the case of an infinite homogeneous plasma. For MHD waves in an inhomogeneous plasma, compression, the three vorticity components, and the three displacement components are all non-zero, and hence the waves have a mix of the properties from waves in a homogeneous plasma. We note that the slow and Alfv\'en resonant waves are driven by the total pressure perturbations $P'$, as pointed out by \citet{Hollweg&Yang1988}.

For waves with a frequency in the cusp continuum, we found that perpendicular vorticity and parallel displacement greatly dominate the other components. Nevertheless, parallel vorticity, though much smaller than perpendicular vorticity, is clearly non-zero in the inhomogeneous layer. These results confirm that the waves have mixed properties, although the waves undergoing resonant absorption in the cusp continuum clearly have properties that closely resemble the properties of the so-called slow waves (i.e. slow magnetosonic waves in a uniform plasma of infinite extent).

For waves with a frequency in the Alfv\'en continuum, it is the parallel component of vorticity and the perpendicular displacement that greatly dominate the other components of the respective quantities. Nevertheless, these other components and quantities, such as compression, are non-zero as well. This again confirms the mixed properties of the waves, which have properties that are closest to the classical Alfv\'en waves when undergoing resonant absorption in the Alfv\'en continuum. This is in line with what was found by \citet{GoossensEtAl2020} for waves undergoing resonant absorption in the Alfv\'en continuum in a pressureless plasma.

\begin{acknowledgements}
 This research was supported by the National Natural Science Foundation of China (BL: 41674172,1111761141002, 41974200; SXC:41604145). M.G. was supported by the C1 Grant TRACEspace of Internal Funds KU Leuven (number C14/19/089). TVD was supported by the European Research
Council (ERC) under the European Union's Horizon 2020 research and
innovation programme (grant agreement No 724326) and the C1 grant
TRACEspace of Internal Funds KU Leuven. The research benefitted greatly
from discussions at ISSI-BJ.
\end{acknowledgements}


\bibliographystyle{aa}
\bibliography{biblio}

%\begin{thebibliography}{}
%
%  \bibitem[Baker(1966)]{baker} Baker, N. 1966,
%      in Stellar Evolution,
%      ed.\ R. F. Stein,\& A. G. W. Cameron
%      (Plenum, New York) 333
%
%   \bibitem[Balluch(1988)]{balluch} Balluch, M. 1988,
%      A\&A, 200, 58
%
%   \bibitem[Cox(1980)]{cox} Cox, J. P. 1980,
%      Theory of Stellar Pulsation
%      (Princeton University Press, Princeton) 165
%
%   \bibitem[Cox(1969)]{cox69} Cox, A. N.,\& Stewart, J. N. 1969,
%      Academia Nauk, Scientific Information 15, 1
%
%   \bibitem[Mizuno(1980)]{mizuno} Mizuno H. 1980,
%      Prog. Theor. Phys., 64, 544
%   
%   \bibitem[Tscharnuter(1987)]{tscharnuter} Tscharnuter W. M. 1987,
%      A\&A, 188, 55
%  
%   \bibitem[Terlevich(1992)]{terlevich} Terlevich, R. 1992, in ASP Conf. Ser. 31, 
%      Relationships between Active Galactic Nuclei and Starburst Galaxies, 
%      ed. A. V. Filippenko, 13
%
%   \bibitem[Yorke(1980a)]{yorke80a} Yorke, H. W. 1980a,
%      A\&A, 86, 286
%
%   \bibitem[Zheng(1997)]{zheng} Zheng, W., Davidsen, A. F., Tytler, D. \& Kriss, G. A.
%      1997, preprint
%\end{thebibliography}

\end{document}